# General Phase Segregation and Phase Pinning Effects in Ln-doped Lead Halide Perovskite with Dual-wavelength Lasing

*Junyu He,[1] Jun Luo,[1] Weihao Zheng,[3,4,*] Biyuan Zheng,[3,4] Mengjian Zhu,[3,4] Jiahao Liu,[3,4] Tingzhao Fu,[3,4] Jing Wu,[2] Zhihong Zhu,[3,4] Fang Wang,[5], Xiujuan Zhuang,[2,*]*

[1]School of Physics and Electronics, Hunan University, Changsha, Hunan 410082, P. R. China

[2]National Key Laboratory of Power Semiconductor and Integration Technology, Engineering Research Center of Advanced Semiconductor Technology and Application of Ministry of Education, College of Semiconductors (College of Integrated Circuits), Hunan University, Changsha 410082, P. R. China

[3]College of Advanced Interdisciplinary Studies & Hunan Provincial Key Laboratory of Novel Nano-optoelectronic Information Materials and Devices, National University of Defense Technology, Changsha, 410073, China.

[4]Nanhu Laser Laboratory, National University of Defense Technology, Changsha 410073, P. R. China.

[5]State Key Laboratory of Infrared Physics, Shanghai Institute of Technical Physics, Chinese Academy of Sciences, 500 Yu Tian Road, Shanghai 200083, P. R. China

[*]Corresponding authors. E-mails: zhengweihao@nudt.edu.cn; zhuangxj@hnu.edu.cn

**Abstract**

Lead halide perovskites (LHPs) exhibit outstanding optoelectronic properties, making them highly promising for applications in various optoelectronics devices. However, rapid ion migration in LHPs not only undermines device stability but also hinders the development of multi-band composite structures, which are crucial to advancing perovskite bandgap engineering and unlocking novel applications. Here, we introduce a novel and general strategy involving both phase segregation and phase pinning by doping $Er^{3+}$ into $CsPb(X_xY_{1-x})_3$ (X, Y = Cl, Br, I) microplates via a simple one-step chemical vapor deposition method. The ion migration is effectively suppressed and a variety of stable multi-band composite structures are demonstrated, with diverse dual-band photoluminescence emissions covering the red, green, and blue spectral bands. The corresponding high-performance dual-wavelength lasers have also been fabricated, confirming the stability and high crystalline quality of these multi-band composite structures. In addition, this strategy is extended to the doping of various lanthanide ion and the incorporation of different mixed halides into LHPs. As a result, a series of multi-band composite structures based on LHPs and corresponding dual-wavelength lasers are developed, thereby validating the generality of this strategy. Theoretical calculations clarify the phase segregation and phase pinning mechanism in these LHPs. Our work not only facilitates the stability design of LHPs but also significantly advances the bandgap engineering, thereby contributing to the expansion of their potential applications in the optoelectronic future.

# Introduction

Lead halide perovskites (LHPs) exhibit outstanding optoelectronic properties, including high absorption coefficients, high photoluminescence (PL) quantum yields, remarkable defect tolerance, tunable bandgaps, and excellent charge transport characteristics[1-4]. These distinctive features have enabled their widespread application in various optical fields, including light-emitting diodes (LEDs)[5-8], photodetectors[9-11], visual displays[12-14], and lasers[15-17]. However, the soft lattice nature of LHPs makes them highly susceptible to rapid halide ion migration when exposed to external stimulation, such as applied electric fields, light irradiation, and temperature changes[18-21]. Such ion migration can induce interfacial halide ion diffusion or segregation, electrode polarization, and hysteresis phenomena[22-24]. These issues not only undermine the stability of the devices but also induce substantial performance degradation. Moreover, ion migration can erase the hetero-interfaces, which presents a significant obstacle to constructing stable multi-band composite structures based on LHPs. How to suppress the ion migration to obtain perovskites both with stability and multi-band composite structures remains a great challenge that urgently needs to be broken through, which are essential for advancing perovskite bandgap engineering and unlocking novel functional applications.

Recently, many studies have been devoted to suppressing ion migration in perovskite materials. Chang *et al.* have significantly suppressed ion migration in perovskite solar cells by generating a highly ordered two-dimensional (2D) perovskite phase on the surface of a 3D perovskite film using a meta-amidinopyridine ligand and a solvent post-dripping step, thereby enhancing the performance and stability of the devices[25]. Wang *et al.* effectively suppressed ion migration by introducing thioacetylacetamide hydrochloride to bind with the cations and anions (named "halide locking") in the perovskite, thereby reducing the formation of ion migration pathways[26]. Interestingly, Hong *et al.* proposed a phase pinning strategy, which stabilizes the disordered phase by suppressing the ordering transition of the spacer cations in two-dimensional perovskites through the introduction of a small amount of pentylammonium or butylammonium as dopants[27].

Drawing on these studies, the potential for achieving ion migration suppression via strategies like doping is vast. In addition, the phase pinning can lead to the suppression of ion migration, which in turn enables the construction of structures with stability. By contrast, phase segregation is a common occurrence in mixed-halide perovskite alloys when subjected to light, heat, or electrical excitation[28]. This process, driven by the migration of halide ions, results in the enrichment of different halide elements and the formation of multi-band composite structures. Regrettably, in the absence of external stimulation, spontaneous ion migration and diffusion typically dismantle these multi-band composite structures, reverting them to single-band alloyed configurations. However, if the phase pinning can be achieved based on the phase segregation, the segregated phase states can be preserved even after the removal of external stimuli, thereby realizing their stability. Such stable phase-separated structures hold great promise for constructing the desired stable multi-band composite structures.

In this work, we direct fabricate Ln-CsPb($X_xY_{1-x}$)$_3$ (X, Y = Cl, Br, I) microplates (MPs) with the initial segregation of different halide phases, by doping with various Ln ion through a simple one-step chemical vapor deposition (CVD) strategy. Representative Er-CsPb($X_xY_{1-x}$)$_3$ MPs are selected for comprehensive and in-depth investigation. Importantly, the halide ion migration in these MPs is effectively suppressed, therefore stabilizing these segregated phases and thus achieving phase pinning of different halides. Based on the simultaneous phase segregation and phase pinning, we achieve the engineered multi-band composite structures with high stability. Stable blue and green dual-band PL emissions in Er-CsPb($Cl_xBr_{1-x}$)$_3$ are observed, as well as green and red, blue and red dual-band PL emissions for Er-CsPb($Br_xI_{1-x}$)$_3$ and Er-CsPb($Cl_xI_{1-x}$)$_3$ MPs, respectively. Additionally, doping with complementary $Yb^{3+}$, $Nd^{3+}$, or $Eu^{3+}$ have similarly achieve both phase segregation and phase pinning effects, demonstrating the generality of this strategy in LHPs. Utilizing these MPs with stable dual-band emissions, we further achieve three types of dual-wavelength lasers with relative low thresholds as well as short lifetime. Such good performances of lasing indicate the good optical features and high-quality of the lattice for cavities in these MPs. Theoretical calculations clarify the phase segregation and pinning mechanism in Ln-doped LHPs. Our finding provide a general and simple strategy for fabricating LHPs with both phase segregation and phase pinning, and eventually demonstrate stable multi-band composite structures. It provides a practical approach to improving the stability of perovskites and significantly expands their application potential in the future optoelectronic field.

**Results and Discussions**

The Er-CsPb($X_xY_{1-x}$)$_3$ MPs were prepared using a precisely controlled CVD method (see Methods and **Fig. S1**). Since the source materials contain two species of halogen elements and a small amount of erbium halide, the prepared samples are usually all-inorganic single-crystal perovskite with three-dimensional structures doped with $Er^{3+}$ in mixed halogen perovskite, as depicted in **Fig. 1**a. When two different halide ions coexist in the perovskites, they typically appear in a homogeneous alloy state, resulting in a single bandgap between that of two pure species, which is a classical method of tuning the bandgap of a compound semiconductor[29]. Therefore, only a single band-edge recombination PL can usually be observed in the mixed halogen perovskites[30]. However, quite an unusual phenomenon was observed in our case. **Fig. 1**b shows the PL spectra and emission photographs (insets in **Fig. 1**b) of the Er-CsPb($Cl_xBr_{1-x}$)$_3$, Er-CsPb($Br_xI_{1-x}$)$_3$, and Er-CsPb($Cl_xI_{1-x}$)$_3$ MPs, where each sample exhibits dual-color emission bands responsible for dual-peak PL shape. The Er-CsPb($Cl_xBr_{1-x}$)$_3$ shows a blue band at 426.1 nm and a green band at 518.6 nm, corresponding to the Cl-rich and Br-rich phase. Similarly, Er-CsPb($Br_xI_{1-x}$)$_3$ shows a green band (Br-rich phase) and a red band (I-rich phase) at 538.3 and 694.3 nm, and the Er-CsPb($Cl_xI_{1-x}$)$_3$ shows a blue band (Cl-rich phase) and a red band (I-rich phase) at 431.8 and 689.7 nm, respectively. Notably, the disappearance luminescence of $Er^{3+}$ in ~520 nm corresponding to $^4I_{15/2}$-$^2H_{11/2}$ transition indicates that $Er^{3+}$ may in the center of the perovskite octahedron because these hypersensitive transitions are forbidden when lanthanide ion are in the central lattice of the Oh group[31].

This Er-doping-induced stable phase segregation fundamentally differs from the photo-, electro-, or thermally-induced phase segregation in undoped LHPs, which has been widely reported previously[32]. The phase segregation observed in our case is not affected by external excitation, including thermal and photoexcitation. As shown in **Fig. 1**c, these dual-band emissions exhibit total stability for their intensities with increasing the number of laser shots. Such significant dual-band emission peaks and their relevant stability differ from the undoped $CsPb(X_xY_{1-x})_3$ MPs, which only show one alloying peak at the initial light excitation time (**Fig. S2**). Based on the above results, we believe that $Er^{3+}$ doping can pin different segregated perovskite phases stably and hinder the halide ion migration in the LHPs. Even more interesting, lasing can be conveniently implemented at each PL band under an optical pump. Each MP is a dual-wavelength semiconductor laser (**Fig. 1**d). Obvious coherent fringes can be observed among these MPs during lasing (insets in **Fig. 1**d).

Confirmatory laser experiments, including laser threshold effect, spectral broadening, and carrier lifetimes, were performed to confirm the above dual-wavelength laser. PL spectra of the representative $Er-CsPb(Cl_xBr_{1-x})_3$ MP under several selected pump fluences are shown in **Fig. 2**a. At low pump fluence, such as 10.1 μJ/cm², the spectrum exhibits two PL peaks with full width at half maximum (FWHM) value of 16~23 nm. While increasing the pump fluence, the sharp PL peak first appears at the Br-rich phase. When increasing the pump fluence to 29 μJ/cm², the Cl-rich phase begins lasing and eventually displays dual-wavelength lasing behavior under the pump fluence of 64.8 μJ/cm². At this time, the FWHM values were narrowed to 1.3~1.8 nm. **Fig. 2**b plots the PL intensity as a function of the pump fluence, which indicates classic *S*-shape curves. **Fig. 2**c and d present their corresponding FWHM values as a function of the pump fluence, respectively. The lasing threshold pump fluences ($P_{th}$) of 15.4 and 29.0 μJ/cm² are obtained for the Cl-rich and Br-rich phase perovskites, respectively. Time-resolved PL (TRPL) spectroscopy is conducted to investigate the PL lifetime performance when lasing. **Fig. 2**e and f give the streak camera images of the Cl-rich and Br-rich phase perovskites from the $Er-CsPb(Cl_xBr_{1-x})_3$ MP, respectively. The bright dots in the two images imply the extremely short PL lifetime when the pump fluence is 1.2 $P_{th}$. After the single exponential fit with deconvolution of the instrument response function (IRF), a time constant is figured out as 3.2 and 2.4 ps for the Cl-rich phase and Br-rich phase perovskites, respectively. Similar dual-wavelength lasing behavior can also be observed in the Er-CsPb(Br$_x$I$_{1-x}$)$_3$ and Er-CsPb(Cl$_x$I$_{1-x}$)$_3$ MPs (**Fig. S3** and **S4**).

TRPL spectroscopy is also employed to investigate the carrier dynamics of the Er-CsPb(X$_x$Y$_{1-x}$)$_3$ MPs before lasing. **Fig. 3**a and b give the steak camera images collected from the PL of the Cl-rich and Br-rich phases of a typical Er-CsPb(Cl$_x$Br$_{1-x}$)$_3$ MP, respectively. The PL decay curves for both phases are extracted from the areas between the white-dashed lines and are shown in **Fig. 3**c. These dynamic curves are fitted by a multi-exponential function with deconvolution of the IRF[33]. The fitting process and results include PL decay time constants and the corresponding amplitudes are listed in **Table S1**. Three pure-phase perovskite samples, CsPbCl$_3$, CsPbBr$_3$, and CsPbI$_3$ MPs, are used to compare and study the carrier dynamics (**Fig. S5** and **Table S2**). Usually, the pure CsPbX$_3$ without Er$^{3+}$ doping only exists two time constants with both positive amplitudes, which are attributed to the carrier recombination through the surface and intrinsic states, respectively[34,35]. However, the PL decay of the Cl-rich phase in the Er-CsPb(Cl$_x$Br$_{1-x}$)$_3$ MP exhibits three time constants with all positive amplitudes, and the rise time of the dynamic curve is close to the IRF. In contrast, the dynamic curve of the Br-rich phase is much longer than the IRF and exhibits a longer lifetime than that in the Cl-rich phase. The Br-rich phase has three time constants with one negative amplitude and two positive values. The time constant with the negative amplitude corresponds to the prolonged rise edge of the dynamic curve, which usually implies an extra significant carrier transfer into the Br-rich phase[36]. The above results demonstrate that there is a band-structure realignment and that there should be a robust energy transfer process between the Cl-rich phase and the Br-rich phase. The Cl-rich phase with a wider bandgap can act as a donor, while the Br-rich phase with a narrower bandgap can act as an acceptor. They form a heterostructure-like system that usually emerges a carrier funneling phenomenon[37], as shown in **Fig. 3**d. Similarly, the carrier funneling phenomenon was also observed in Er-CsPb(Br$_x$I$_{1-x}$)$_3$ (**Fig. S6**, **Table S3**) and Er-CsPb(Cl$_x$I$_{1-x}$)$_3$ MPs. (**Fig. S7**, **Table S4**) MPs. The TRPL confirms the phase segregation phenomenon from the bandgap realignment caused by the mixed halide dealloying.

According to the emission peak position, the specific values of the components in each phase can be defined through the conventional bowing equation: $E_{g(x)}=xE_{g1}+(1-x)E_{g2}-mx(1-x)$[38]. The Er-CsPb(Cl$_x$Br$_{1-x}$)$_3$ is calculated with Cl-rich phase (Er-CsPb(Cl$_{0.9}$Br$_{0.1}$)$_3$) and Br-rich phase (Er-CsPb(Cl$_{0.8}$Br$_{0.2}$)$_3$); Er-CsPb(Br$_x$I$_{1-x}$)$_3$ are calculated with Br-rich phase (Er-CsPb(Br$_{0.97}$I$_{0.03}$)$_3$) and I-rich phase (Er-CsPb(Br$_{0.27}$I$_{0.73}$)$_3$), and Er-CsPb(Cl$_x$I$_{1-x}$)$_3$ are calculated with Cl-rich phase (Er-CsPb(Cl$_{0.9}$I$_{0.1}$)$_3$) and I-rich phase (Er-CsPb(Cl$_{0.1}$I$_{0.9}$)$_3$) (the calculation procedure is shown in SI). According to the calculated results, the above blue, green, and red bands are close to the band-edge PL of pure phase perovskite of CsPbCl$_3$, CsPbBr$_3$, and CsPbI$_3$, respectively (**Fig. S8**). For example, in Er-CsPb(Cl$_x$Br$_{1-x}$)$_3$, the atom ratio of Cl:Br in the Cl-rich phase reaches 9:1, while the atom ratio of Br:Cl in Br-rich phase reaches 8:1. Why does the usually alloyed system of the mixed halide perovskites exhibit a significant phase segregation performance? Materials characterization was first performed to clarify the mechanism. From the corresponding optical images of the Er-CsPb(Cl$_x$Br$_{1-x}$)$_3$, Er-CsPb(Br$_x$I$_{1-x}$)$_3$, and Er-CsPb(Cl$_x$I$_{1-x}$)$_3$ MPs (**Fig. 4**a-c), it can be seen that three MPs all feature a regular and crystalline morphology, showing the equilateral square shape with the side lengths of several micrometers. From the scanning electron microscopy (SEM) shown in **Fig. 4**d, most samples exhibit good shape, confirming the preferable homogeneity during the growth process. **Fig. 4**e presents the high-magnification SEM image for a representative single Er-CsPb(Cl$_x$Br$_{1-x}$)$_3$ MP, which shows a well-defined morphology without obvious surface defects. Corresponding element mapping images are shown in **Fig. 4**f$_1$-f$_5$, which demonstrate that Cs, Pb, Cl, Br, and Er elements are homogeneously distributed over the entire MP. The energy dispersive X-ray (EDX) results indicates that the Er elements accounted for 3.1% of the total element content in Er-CsPb(Cl$_x$Br$_{1-x}$)$_3$ MP. **Fig. S9** gives similar results for Er-CsPb(Br$_x$I$_{1-x}$)$_3$ and Er-CsPb(Cl$_x$I$_{1-x}$)$_3$ MPs, where the Er element concentration are 3.3 % and 4.4 % of the total element content, respectively. The atomic ratio of Cs:(Pb+Er):(Cl+Br) is about 1:0.91:3, close to the quantitative ratio of $CsPbX_3$ perovskite, which is consistent with the study that the $Er^{3+}$ doped into the lattice are more accessible to replace the B-site ions, $Pb^{2+}$.[39] Based on the EDX results, the chemical formula of three selected MPs can be given as Er-$CsPb(Cl_{0.83}Br_{0.17})_3$, Er-$CsPb(Br_{0.81}I_{0.19})_3$, and Er-$CsPb(Cl_{0.83}I_{0.17})_3$, respectively. **Fig. 4**g shows the X-ray diffraction (XRD) patterns of the Er-$CsPb(Cl_{0.83}Br_{0.17})_3$ MPs compared with the undoped $CsPb(Cl_{0.83}Br_{0.17})_3$ MPs. Two kinds of MPs possess diffraction peaks that match with the (100), (110), and (200) crystal planes of the cubic $CsPbX_3$ lattice structure, which indicates that $Er^{3+}$ doping does not change the lattice structure of the perovskite. Compared to the undoped sample, the prominent diffraction peak of Er-$CsPb(Cl_{0.83}Br_{0.17})_3$ shifted to a larger angle apparently, which is caused by lattice shrink when small-size $Er^{3+}$ (ion radius: 0.88 Å) replacing large-size $Pb^{2+}$ (ion radius: 1.2 Å), as schematically shown in **Fig. 4**h. Both the EDX and the XRD results have proved that $Er^{3+}$ is successfully doped into the lattice of the host perovskite. This intrinsic phase segregation and phase pinning is not limited to the case of $Er^{3+}$ doping alone. It is also observed in the doping of various other lanthanide ion such as the doping of $Yb^{3+}$, $Nd^{3+}$, and $Eu^{3+}$ as shown in **Fig. S10**. The effects induced by the trivalent lanthanide ion doping are highly similar, indicating that the mechanisms of heterovalent substitution of Pb and the consequent phase segregation and phase pinning are analogous.

In order to investigate the specific mechanism of $Er^{3+}$ doping promoting phase segregation and phase pinning, we carried out first-principle calculations and analyzed the effect of $Er^{3+}$ from the perspective of formation energy and ion migration. To clarify the types of defects that doping may induce, we first calculate the formation energies of different defect types in Er-$CsPb(Cl_xBr_{1-x})_3$. According to **Fig. 5**a, only the defect in which $Er^{3+}$ substitutes $Pb^{2+}$ exhibits a negative formation energy, indicating that this defect is the most likely to form. Therefore, we focused on how this type of defect influences phase segregation. Next, we calculated the formation energies of two lattice structural configurations, the homogeneous alloy distribution case and the regional halogen aggregation case, for both undoped $CsPb(Cl_{0.83}Br_{0.17})_3$ (averagely containing one Br per octahedron) and Er- CsPb(Cl$_{0.83}$Br$_{0.17}$)$_3$ perovskites (**Fig. 5**b, **Table S5**). The results show that the formation energy of local halogen aggregation distribution is 0.1 eV higher than that of homogeneously mixed halogen distribution in CsPb(Cl$_{0.83}$Br$_{0.17}$)$_3$ For Er-doped sample, the formation energy of the local halogen aggregation case is 0.97 eV lower than that of the homogeneously mixed halogen distribution case. The significant energy difference indicates that Er$^{3+}$ doping makes the local halogen aggregation case more stable than the homogeneously mixed halogen distribution case, which strongly promotes the intrinsic phase segregation phenomenon. The mechanism can be understood in the following way: when one halogen species aggregates, the local chemical potential and electronegativity in this region will change significantly. These variations will lead to a more significant charge density gradient, which makes the interaction between ions uneven and further intensifies the chemical instability of adjacent regions, prompting the aggregates region to separate from the other areas and form pinning phase segregation.

The ion migration barrier represents a fundamental criterion for assessing the migration propensity of halogen ions. Since the ratio of Cl and Br in this case is about 5:1, the migration capacity of Cl$^-$ is closely related to phase segregation. Based on this, we further calculated the intra- and inter-octahedral migration barriers of Cl$^-$ before and after Er$^{3+}$ doping (**Fig. 5**c and, **Table S6**). After Er$^{3+}$ doping, a notable increase in Cl$^-$ migration barriers is observed. Specifically, the intra-octahedral migration barrier elevates from 0.56 eV to 1.58 eV after Er$^{3+}$ doping, with a corresponding increase in the inter-octahedral migration barrier from 1.00 eV to 1.83 eV. Overall, these findings demonstrate that Er$^{3+}$ doping facilitates the formation of separated perovskite phases by phase pinning strategy, and also effectively suppresses the ion migration to improve the stability of system (**Fig. 5**d).

In summary, an innovative strategy that simultaneously utilizes phase segregation and phase pinning effects to achieve stable multi-band composite structures in Er-CsPb(X$_x$Y$_{1-x}$)$_3$ MPs is demonstrated. Through a simple one-pot CVD method, three stable multi-band composite structures were fabricated, which exhibit three types of dual-band PL emission spanning the red, green, and blue spectral regions with stable photo-excitation stability. Furthermore, we have realized three types of dual-wavelength lasers with thresholds in the tens of μJ/cm$^2$ and lifetimes on the order of several picoseconds. These high-quality dual-wavelength lasers, in turn, demonstrate that the prepared multi-band composite structures possess high-quality lattice structures and high stability. TRPL spectra experiments manifest a robust energy transfer within these separated phases. From the perspective of forming energy and ion migration through first-principles calculation, lanthanum ions doping makes the local halogen aggregation more stable than the homogeneously mixed halogen distribution case, which is conducive to local halogen aggregation (phase segregation) and phase pinning. In addition, the Er$^{3+}$ doping also increases the migration barrier of Cl$^-$ and inhibits ion migration. This work develops a brand-new strategy for controlling halide ions migration and constructing multi-band composite structures with high stability and lattice quality for LHPs. This advancement significantly broadens the application scope of LHPs in photonics and optoelectronics.

**METHODs**

**Sample Preparation** Er-CsPb($X_xY_{1-x}$)$_3$ MPs were grown using a CVD method. For Er-CsPb(Cl$_x$Br$_{1-x}$)$_3$ MPs, ErCl$_3$ (99.999%, Alfa), CsBr (99.999%, Alfa), PbCl$_2$ (99.999%, Alfa) powers are mixed in a clean alumina boat with a mass ratio of 3:1:1 and then placed in the heating zone of a tubular furnace. For Er-CsPb(Br$_x$I$_{1-x}$)$_3$ MPs, ErBr$_3$ (99.999%, Alfa), CsI (99.999%, Alfa), PbBr$_2$ (99.999%, Alfa) powers are mixed in a clean alumina boat with a mass ratio of 6:1:3 and then placed in the heating zone of a tubular furnace. For Er-CsPb(Cl$_x$I$_{1-x}$)$_3$ MPs, ErCl$_3$ (99.999%, Alfa), CsI (99.999%, Alfa), PbCl$_2$ (99.999%, Alfa) powers are mixed in a clean alumina boat with a mass ratio of 3:1:1 and then placed in the heating zone of a tubular furnace. The freshly cleaved fluorophlogopite mica is positioned downstream of the furnace as the substrate. To clean the impurity gas, a high flow rate of carrier gas of Ar is maintained for 10 minutes (>500 sccm). Then, the pressure of the system is decreased to less than 20 mm Torr by a vacuum pump and a constant flow of ~20 sccm for the carrier gas during the whole growth. The corresponding thermodynamic parameters were set according to the different target products: for Er-CsPb(Cl$_x$Br$_{1-x}$)$_3$, the temperature was raised to 820 °C and held for 6 minutes; For Er-CsPb(Br$_x$I$_{1-x}$)$_3$, the temperature was raised to 960 °C and held for 14 minutes; For Er-CsPb(Cl$_x$I$_{1-x}$)$_3$, the temperature is heated to 830 °C and held for 6 minutes. The whole system is cooled down naturally after the growth process.

**Optical Measurements** Steady-state PL spectra and TRPL measurements were detected based on a confocal microscope system (WITec, alpha-300). A mode-locked Ti: sapphire laser (Tsunami) at 800 nm (pulse width 80 fs, repetition frequency 80 MHz) is introduce to a BBO crystal to obtain a second harmonic generation with 400 nm wavelength as the excitation light source. Dark-field image measurements were conducted by introducing the corresponding laser through the 50-fold lens to excite the MPs without any other illumination. For lasing experiments, the same confocal microscope was employed. The 800 nm laser was amplified by a regenerative amplifier (Spitfire Ace 100, 1 kHz) and passed through a BBO crystal to produce 400 nm light (1 kHz) for excitation. The lasing beam was focused by an Olympus 20× objective lens to create a uniform spot with a diameter of 200 μm on the sample. The PL and lasing signals were detected by a spectrometer. While TRPL experiments of both PL and lasing were measured by reflecting the light signal into the streak camera (C10910, Hamamatsu) by Ag mirrors.

**Theoretical Calculations** Density functional theory (DFT) calculations are implemented by the Vienna ab initio Simulation Package (VASP). In this study, the projected augmented wave (PAW) method was used to describe the interaction between valence electrons and ionic nuclei. The energy cutoff of the plane wave base group is set to 300eV. The convergence standard of the force is 0.02 eV/ A. All calculations of the defect structure, i.e., formation energy, are performed in 3×3×3 supercells, so the K-point grid is set to 1×1×1. Material formation energy $E_f$ refers to the change in energy that introduces atoms from the standard state of the element into the material under study. In the calculation the formation of the material can be calculated by the following formula:

$$E_f = E_{total} - \sum_i n_i E_i^{element}$$

Where $E_{total}$ is the total energy of the material, $E_i^{element}$ is the energy of the $i_{th}$ element in its standard state, and $n_i$ is the number of moles of the $i_{th}$ component in the material. The total energy $E_{total}$ of the material is calculated by structural optimization, and the elemental quality of the corresponding element calculates the elemental energy of the standard state.

A migration barrier, $E_b$, is a barrier that describes the energy required for an ion or atom to move from one location to another, which can be calculated using the following formula:

$$E_b = E_{TS} - \frac{E_{start} + E_{end}}{2}$$

Where $E_{TS}$ is the energy of the transition state, and $E_{start}$ and $E_{end}$ are the energy of the initial state (starting positions) and final state (target positions), respectively. The energy of the transition state is searched by linearly inserting multiple intermediate configurations between the starting and target locations.

AUTHOR INFORMATION

Corresponding Author

Weihao Zheng, College of Advanced Interdisciplinary Studies, National University of Defense Technology, China

E-mail: zhengweihao@nudt.edu.cn

Xiujuan Zhuang, College of Semiconductors (College of Integrated Circuits), Hunan University, China

Email: zhuangxj@hnu.edu.cn

**Author Contributions**

Junyu He carried out the conception of experimental ideas, the analysis of data, and the writing of the manuscript; Luo Jun carried out theoretical calculations and data processing; Weihao Zheng completed the verification of experimental data, the provision of ideas, and the revision of the manuscript; Biyuan Zheng, Mengjian Zhu and Jiahao Liu performed the optical characterizations; Tingzhao Fu, Jing Wu and Zhihong Zhu completed the preparation of experimental reagents; Fang Wang assisted in the analysis of experimental data and the revision of the manuscript; Xiujuan Zhuang carried out the conceptualization, the formal analysis and the funding acquisition. All authors discussed the results and commented on the manuscript.

Notes

The authors declare no competing financial interest.

**Acknowledgments**

The research is supported by the "111 Center" (B25033) and the National Natural Science Foundation of China (Nos. 61635001, 52002125).

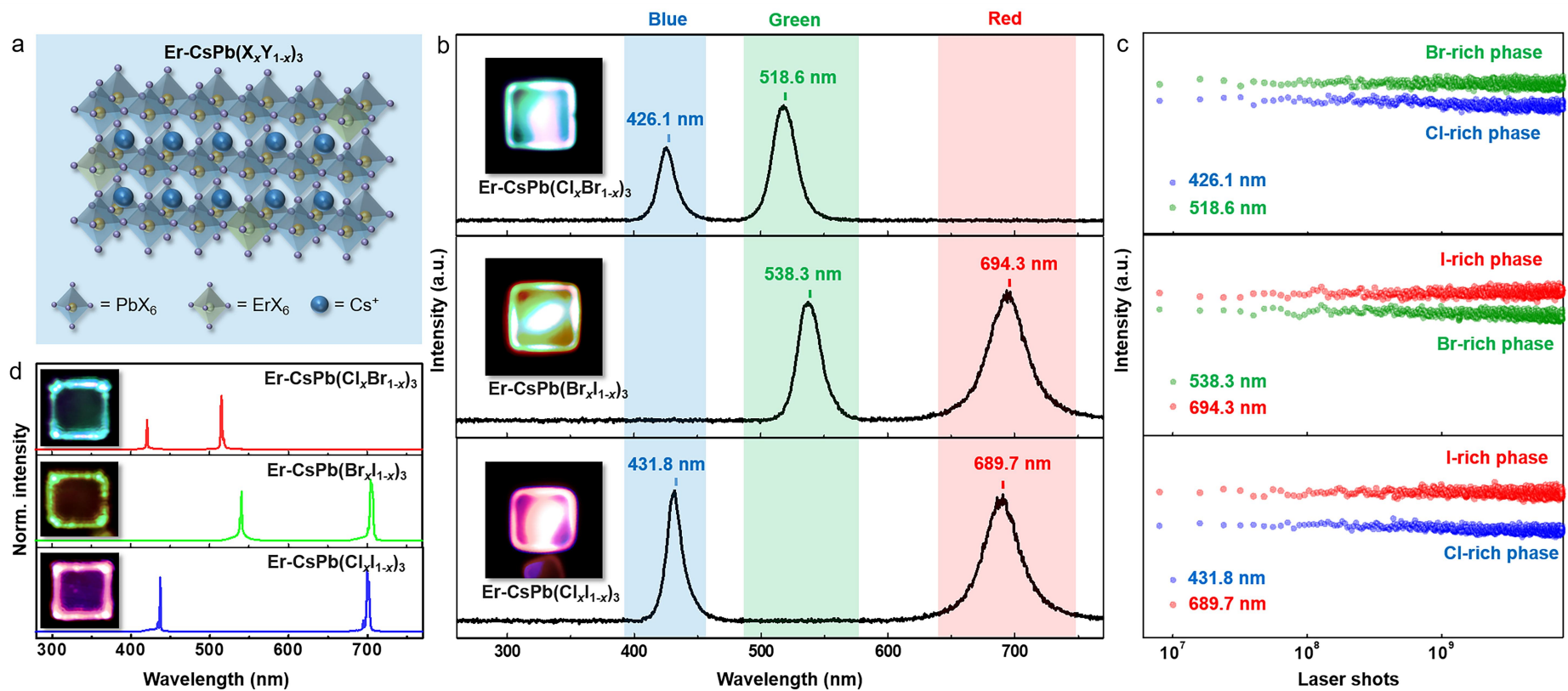

**Fig.1 | Optical characterization of Er-CsPb($X_xY_{1-x}$)$_3$ MPs. a**, lattice structure diagram of Er-CsPb($X_xY_{1-x}$)$_3$ MPs. **b,** PL spectra of Er-CsPb($Cl_xBr_{1-x}$)$_3$, Er-CsPb($Br_xI_{1-x}$)$_3$ and Er-CsPb($Cl_xI_{1-x}$)$_3$ MPs, respectively. The insets are the corresponding PL photographs. **c,** Corresponding emission intensities of each phase in three MPs as a function of the exciting laser shots. **d**, PL spectra of Er-CsPb($Cl_xBr_{1-x}$)$_3$, Er-CsPb($Br_xI_{1-x}$)$_3$ and Er-CsPb($Cl_xI_{1-x}$)$_3$ MPs after lasing. The insets are the corresponding photographs during lasing.

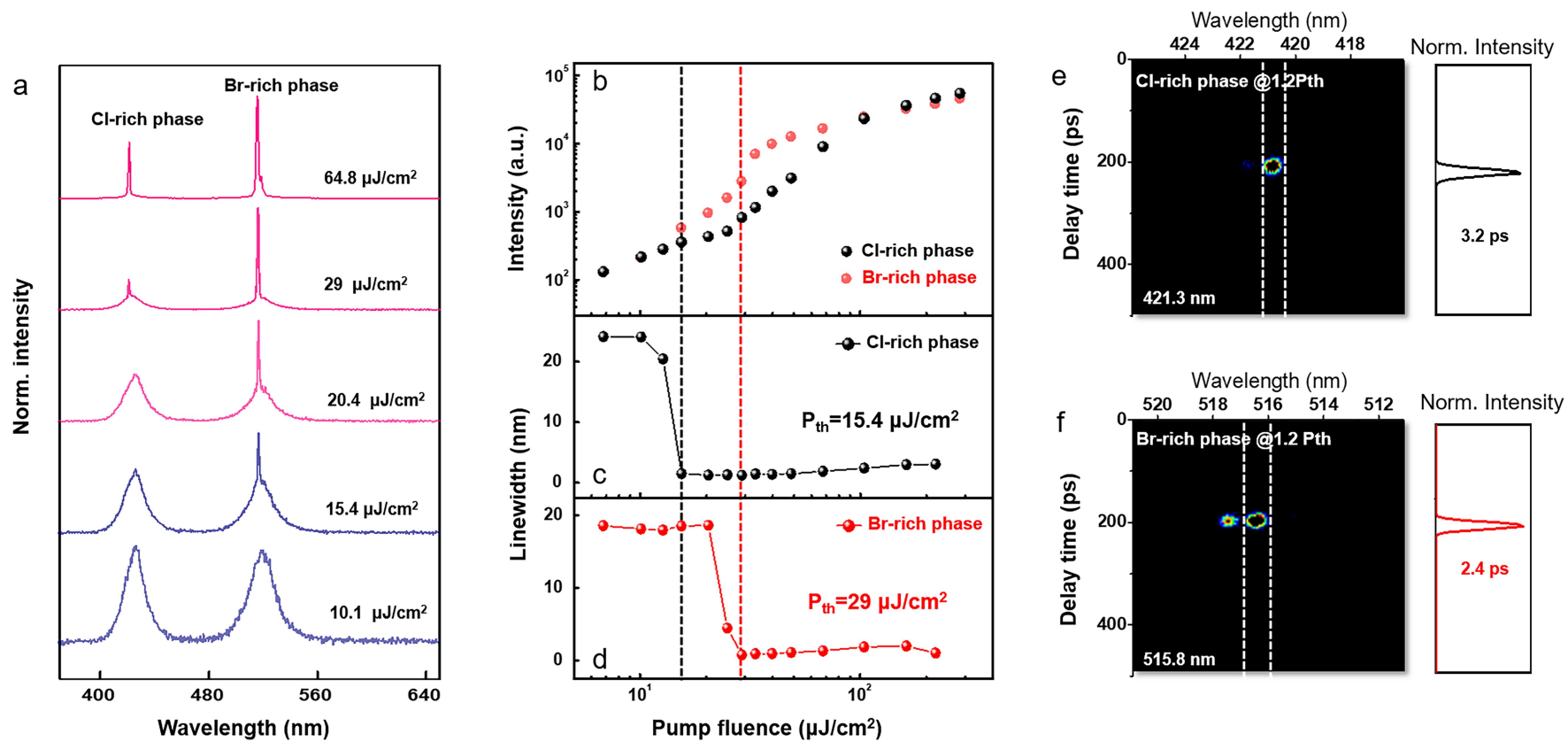

**Fig.2 | Lasing behavior characterization of Er-CsPb(Cl$_x$Br$_{1-x}$)$_3$ MPs. a,** Representative lasing evolution of the Er-CsPb(Cl$_x$Br$_{1-x}$)$_3$ MP under several selected pump fluences. **b,** PL intensities as a function of the pump fluence for both Cl-rich and Br-rich phases. **c,** Corresponding FWHM statistics of Cl-rich as a function of the pump fluence. **d,** FWHM statistics of Br-rich phase as a function of the pump fluence. **e,** Steak camera image of Cl-rich after lasing and the corresponding dynamic curve.
**f,** Steak camera image of Br-rich phase after lasing and the corresponding dynamic curve.

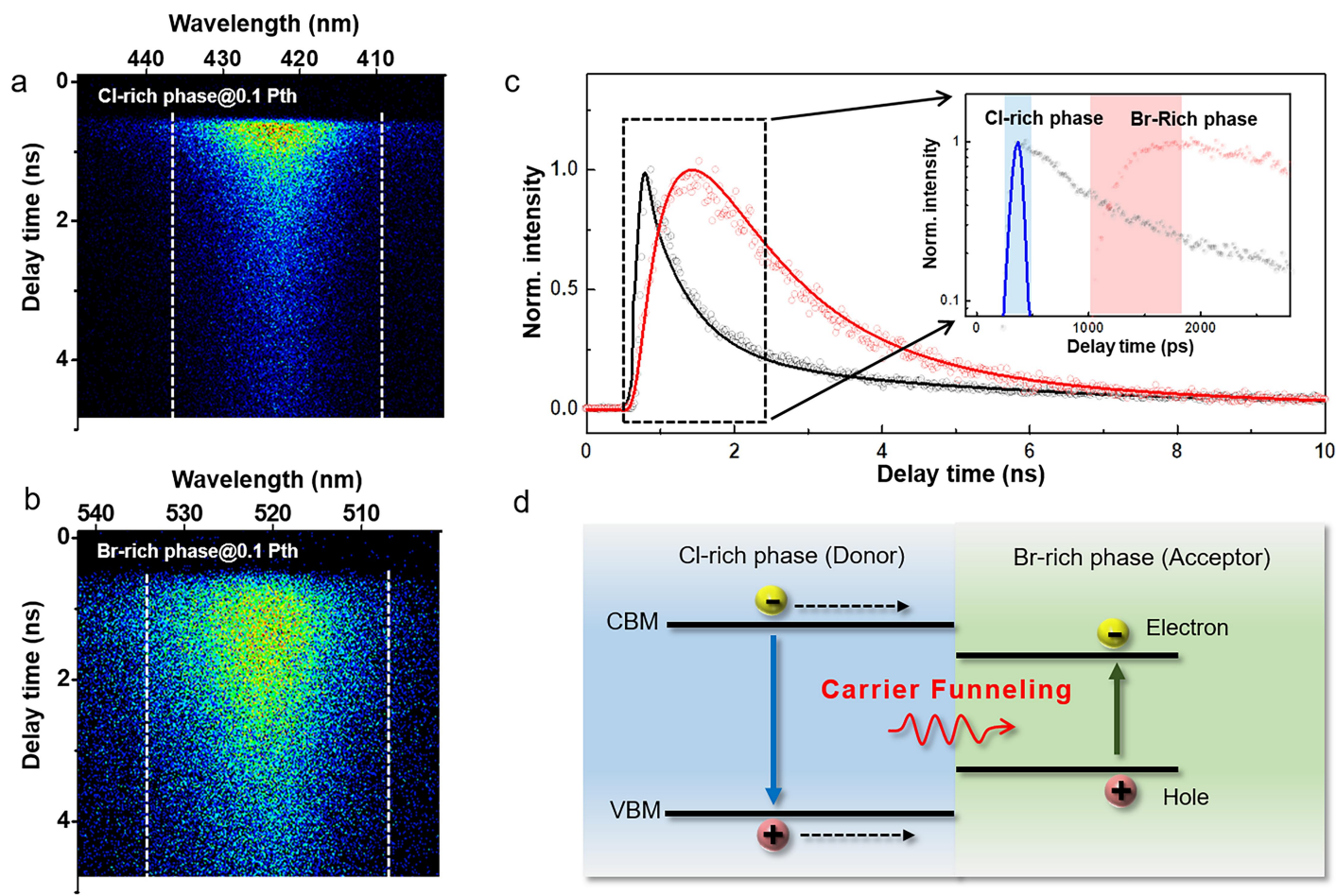

**Fig.3 | Transient spectral dynamic of the representative Er-CsPb(Cl$_x$Br$_{1-x}$)$_3$ MP. a**, Steak camera images collected in the Cl-rich phase emission band of Er-CsPb(Cl$_x$Br$_{1-x}$)$_3$ MP. **b,** Steak camera images collected in the Br-rich phase emission band of Er-CsPb(Cl$_x$Br$_{1-x}$)$_3$ MP. **c,** PL decay curves of Cl-rich phase and Br-rich phase perovskites of Er-CsPb(Cl$_x$Br$_{1-x}$)$_3$ MP. **d,** Schematic diagram of the energy transfer mechanism in the Er-CsPb(Cl$_x$Br$_{1-x}$)$_3$ system.

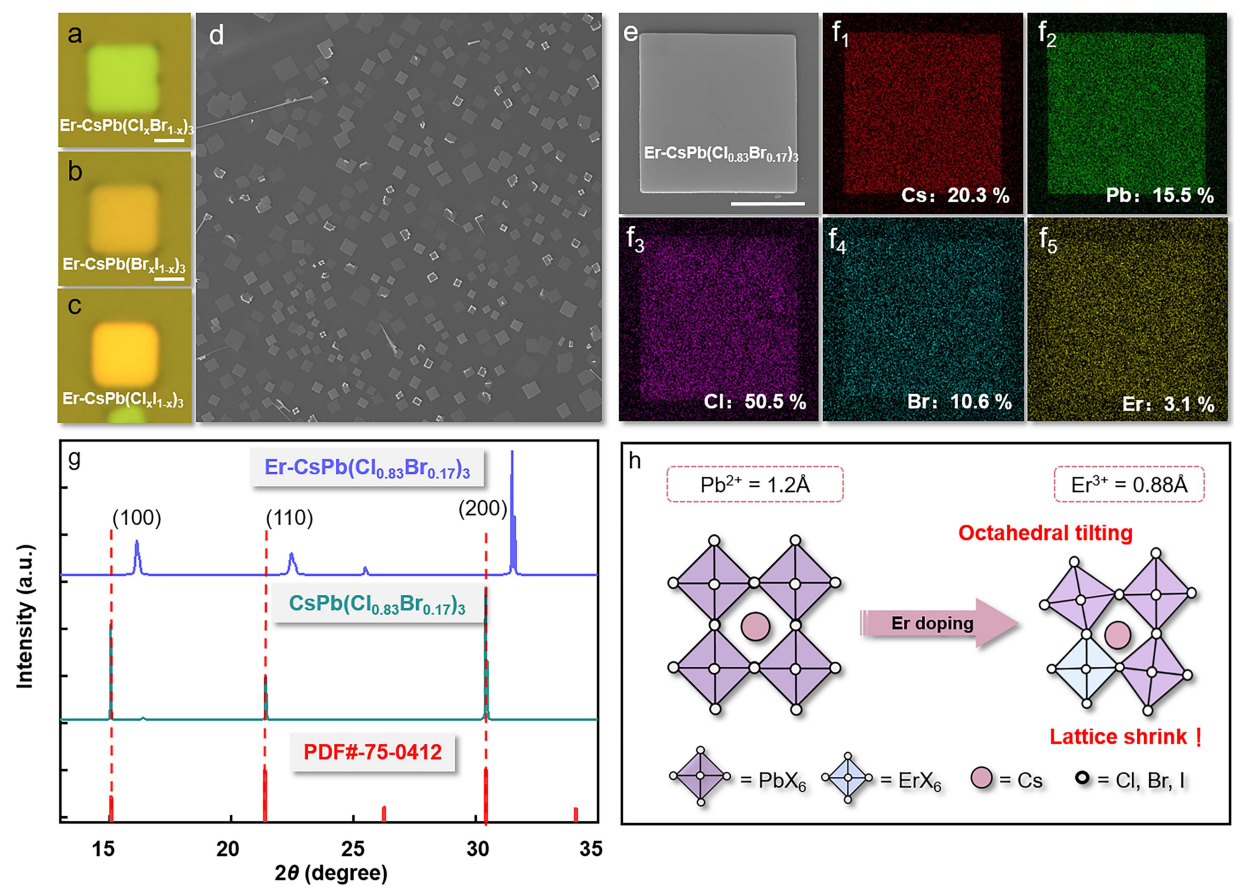

**Fig.4 | Morphology and structure characterization of Er-CsPb($X_xY_{1-x}$)$_3$ MPs. a-c,** The optical images of Er-CsPb($Cl_xBr_{1-x}$)$_3$, Er-CsPb($Br_xI_{1-x}$)$_3$ and Er-CsPb($Cl_xI_{1-x}$)$_3$, respectively. Scale bar: 2 μm. **d,** Large-area low-magnification SEM image of typical Er-CsPb($Cl_xBr_{1-x}$)$_3$ MPs. **e,** High-magnification SEM image of a selected Er-CsPb($Cl_xBr_{1-x}$)$_3$ MP. Scale bar: 2 μm. **f$_1$-f$_5$,** Element mapping of the typical Er-CsPb($Cl_{0.83}Br_{0.17}$)$_3$ MP. **g,** XDR patterns of CsPb($Cl_{0.83}Br_{0.17}$)$_3$ and Er-CsPb($Cl_{0.83}Br_{0.17}$)$_3$ MPs. **h,** Lattice structure diagram before and after Er$^{3+}$ doping.

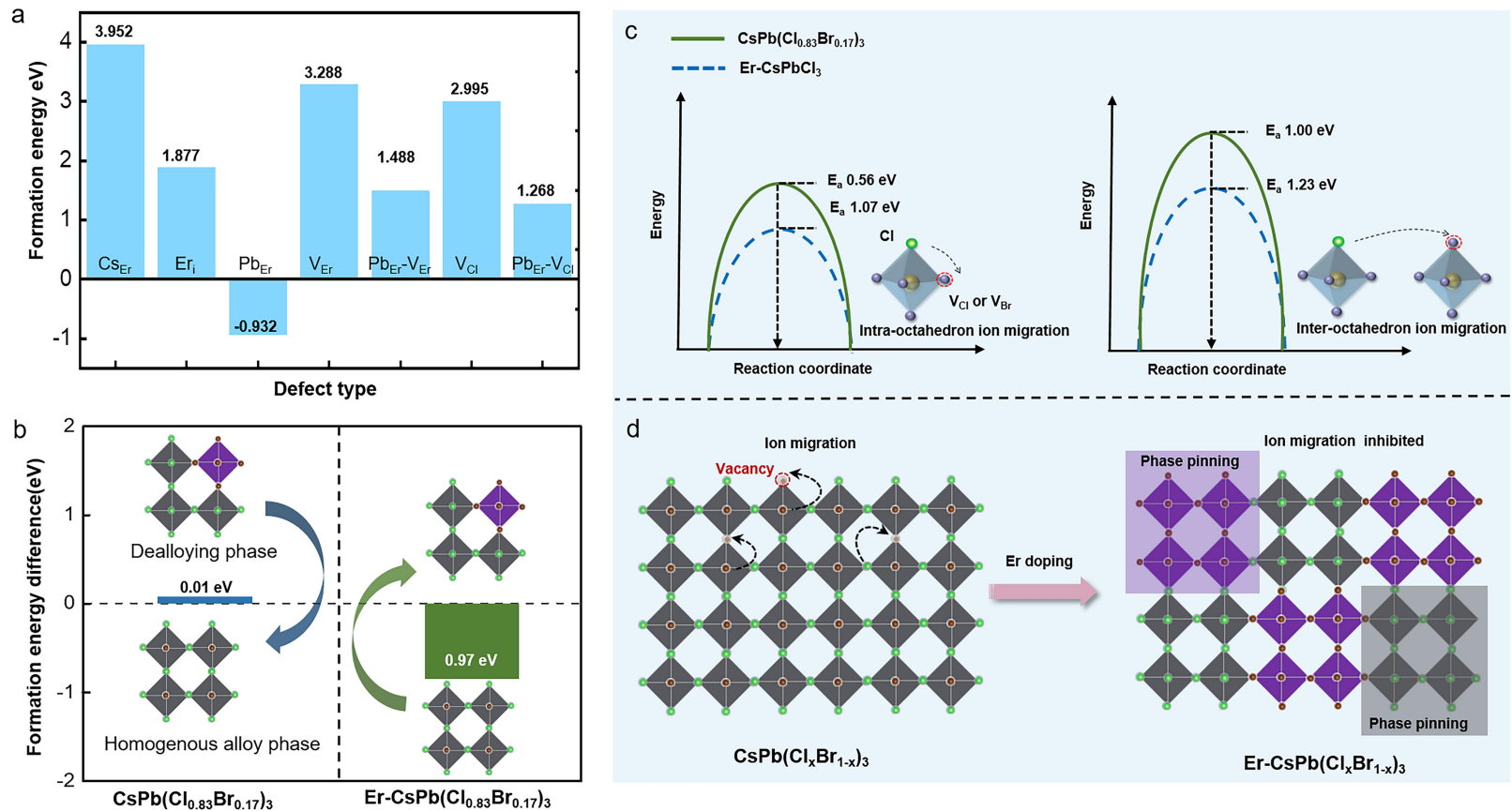

**Fig.5 | Theoretical calculation of the formation energy and ion migration barrier. a,** Formation energy chart of different defects in Er-doping system. **b,** Diagram of formation energy of $CsPb(Cl_{0.83}Br_{0.17})_3$ and $Er$-$CsPb(Cl_{0.83}Br_{0.17})_3$ in the homogeneous alloy distribution case and the regional halogen aggregation case, respectively. **c,** Intra- and inter-octahedral migration barrier calculation of Cl ions in $CsPb(Cl_{0.83}Br_{0.17})_3$ and $Er$-$CsPbCl_3$. **d,** Schematic diagram of $Er^{3+}$ doping promoted the formation of aggregation phase and inhibited ion migration.

# Supporting Information

# General Phase Segregation and Phase Pinning Effects in Er-doped Lead Halide Perovskite with Dual-wavelength Lasing

*Junyu He,[1] Jun Luo,[1] Weihao Zheng,[3,4,*] Biyuan Zheng,[3,4] Mengjian Zhu,[3,4] Jiahao Liu,[3,4] Tingzhao Fu,[3,4] Jing Wu,[2] Zhihong Zhu,[3,4] Fang Wang,[5] Xiujuan Zhuang,[2,*]*

[1]School of Physics and Electronics, Hunan University, Changsha, Hunan 410082, P. R. China

[2]National Key Laboratory of Power Semiconductor and Integration Technology, Engineering Research Center of Advanced Semiconductor Technology and Application of Ministry of Education, College of Semiconductors (College of Integrated Circuits), Hunan University, Changsha 410082, P. R. China

[3]College of Advanced Interdisciplinary Studies & Hunan Provincial Key Laboratory of Novel Nano-optoelectronic Information Materials and Devices, National University of Defense Technology, Changsha, 410073, China.

[4]Nanhu Laser Laboratory, National University of Defense Technology, Changsha 410073, P. R. China.

[5]State Key Laboratory of Infrared Physics, Shanghai Institute of Technical Physics, Chinese Academy of Sciences, 500 Yu Tian Road, Shanghai 200083, P. R. China

[*]Corresponding authors. E-mails: zhengweihao@nudt.edu.cn; zhuangxj@hnu.edu.cn

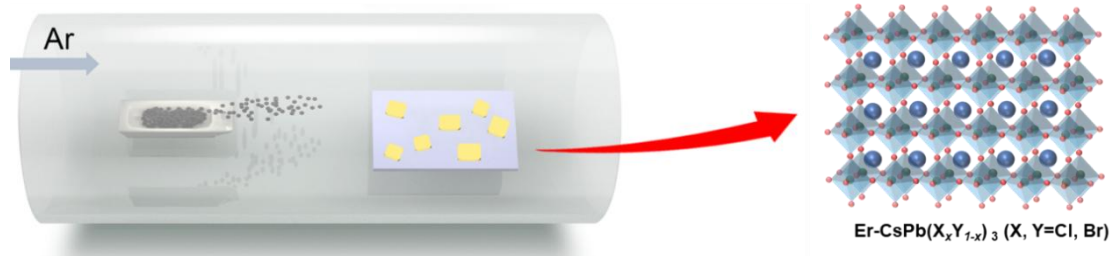

**Fig. S1 | Schematic diagram of Er-CsPb($X_xY_{1-x}$)$_3$ MPs prepared by CVD method.**

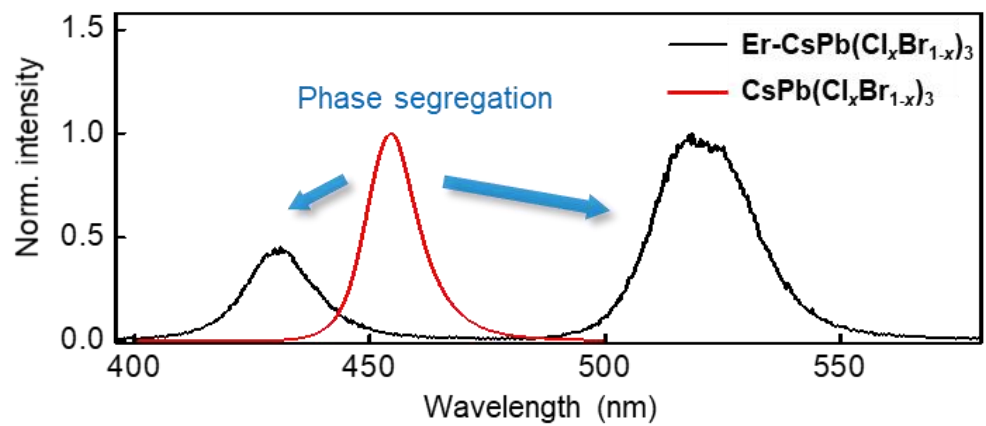

**Fig. S2 | Steady PL spectra of CsPb(Cl$_x$Br$_{1-x}$)$_3$ and Er-CsPb(Cl$_x$Br$_{1-x}$)$_3$ MPs at initial photoexcitation.**

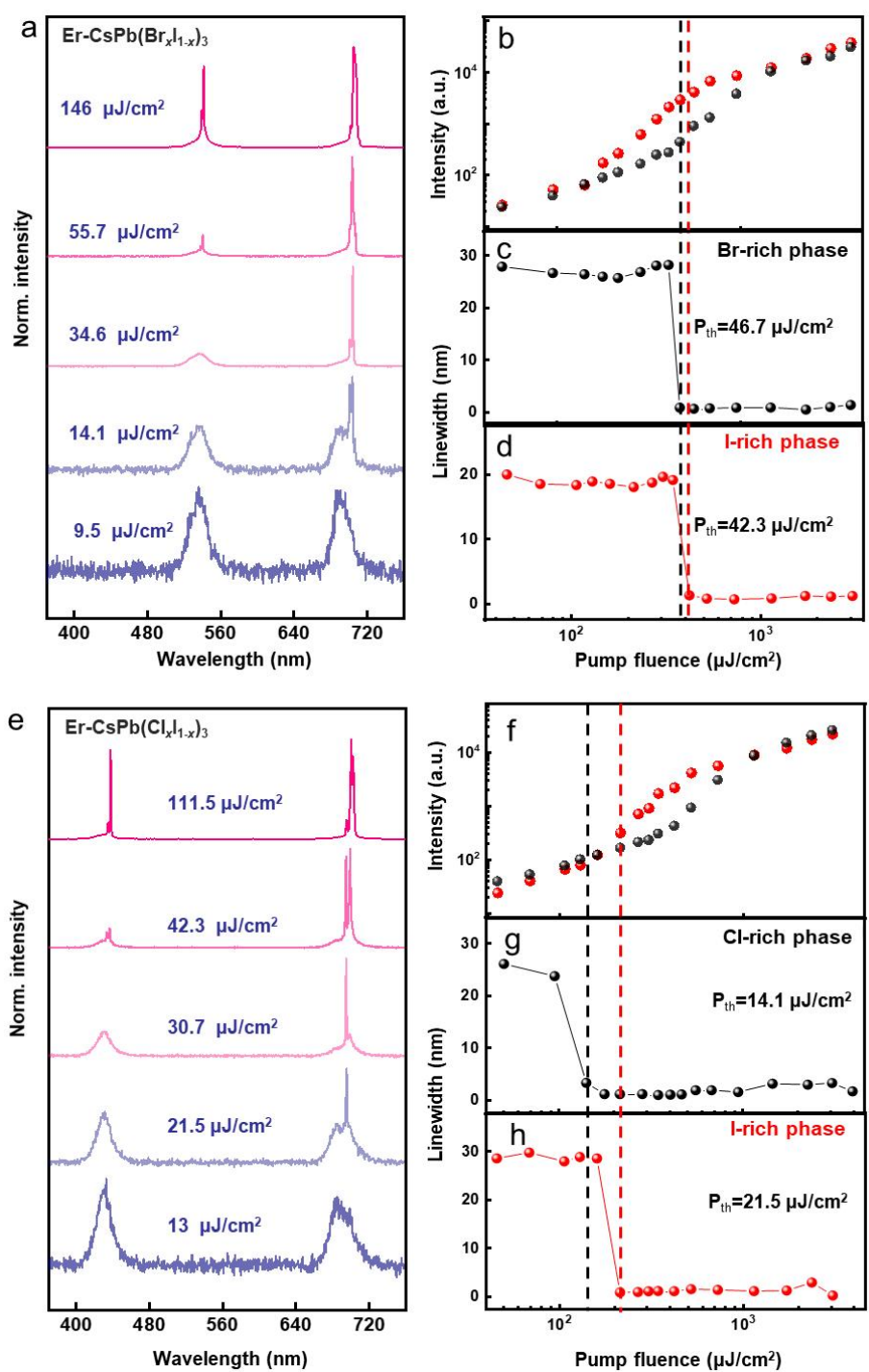

**Fig. S3 | Lasing behavior characterizations of Er-CsPb(Br$_x$I$_{1-x}$)$_3$ and Er-CsPb(Cl$_x$I$_{1-x}$)$_3$ MPs. a,** The pump power-dependent PL spectra of the Er-CsPb(Br$_x$I$_{1-x}$)$_3$ MP. **b,** PL intensities and FWHM of **c,** Br-rich and **d,** I-rich phase as a function of the pump fluence. **e,** The pump power-dependent PL spectra of the Er-CsPb(Cl$_x$I$_{1-x}$)$_3$ MP. **f,** PL intensities and FWHM of **g,** Cl-rich and **h,** I-rich phase as a function of the pump fluence.

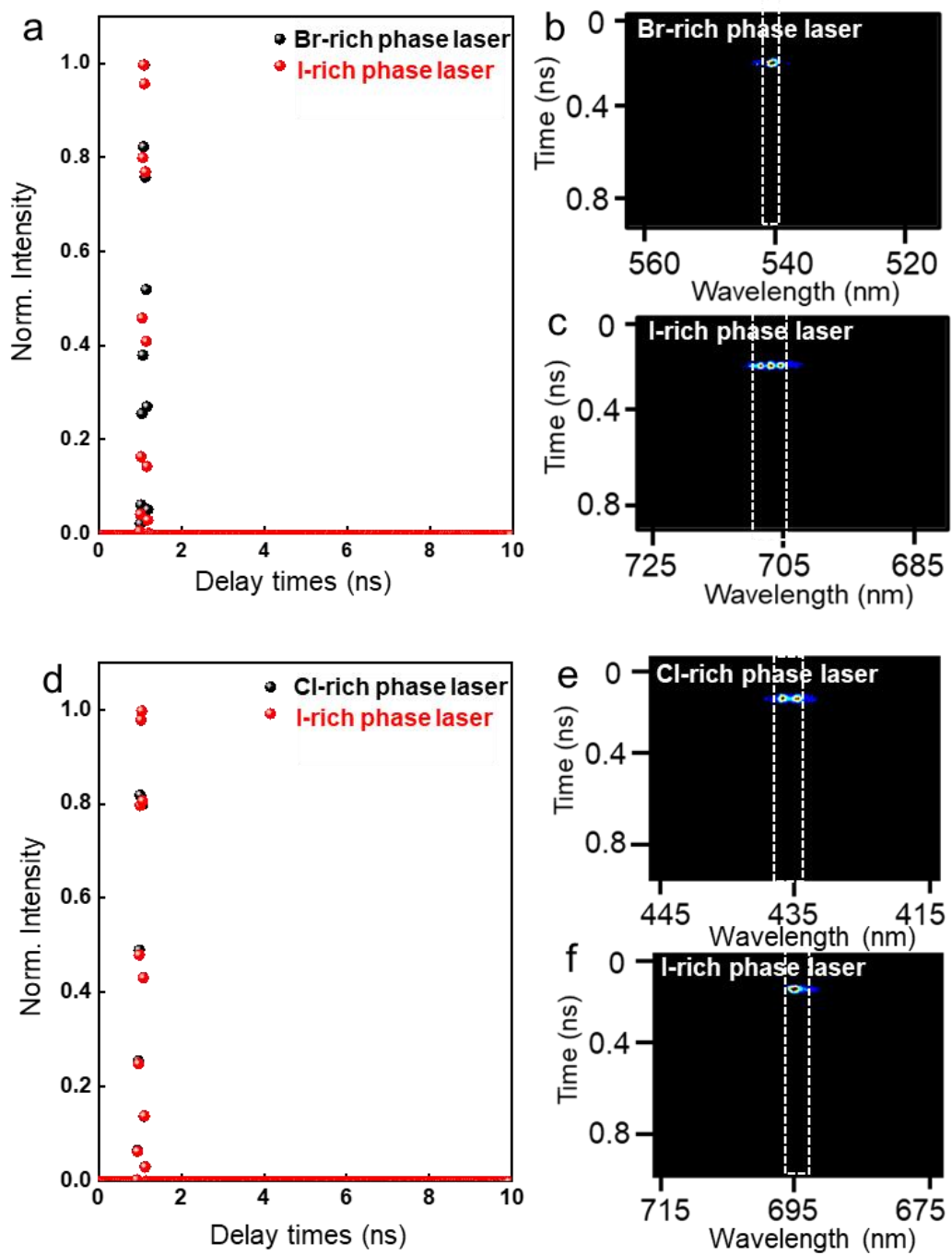

**Fig. S4 | Transient dynamic characterization of the representative Er-CsPb(Br$_x$I$_{1-x}$)$_3$ and Er-CsPb(Cl$_x$I$_{1-x}$)$_3$ MPs after lasing. a,** PL decay curves of Br-rich phase and I-rich phase perovskite in Er-CsPb(Cl$_x$I$_{1-x}$)$_3$. Steak camera image of **b,** Br-rich phase and **c,** I-rich phase after lasing. **d,** PL decay curves of Cl-rich phase and I-rich phase perovskite in Er-CsPb(Cl$_x$I$_{1-x}$)$_3$. Steak camera image of **e,** Cl-rich phase and **f,** I-rich phase after lasing.

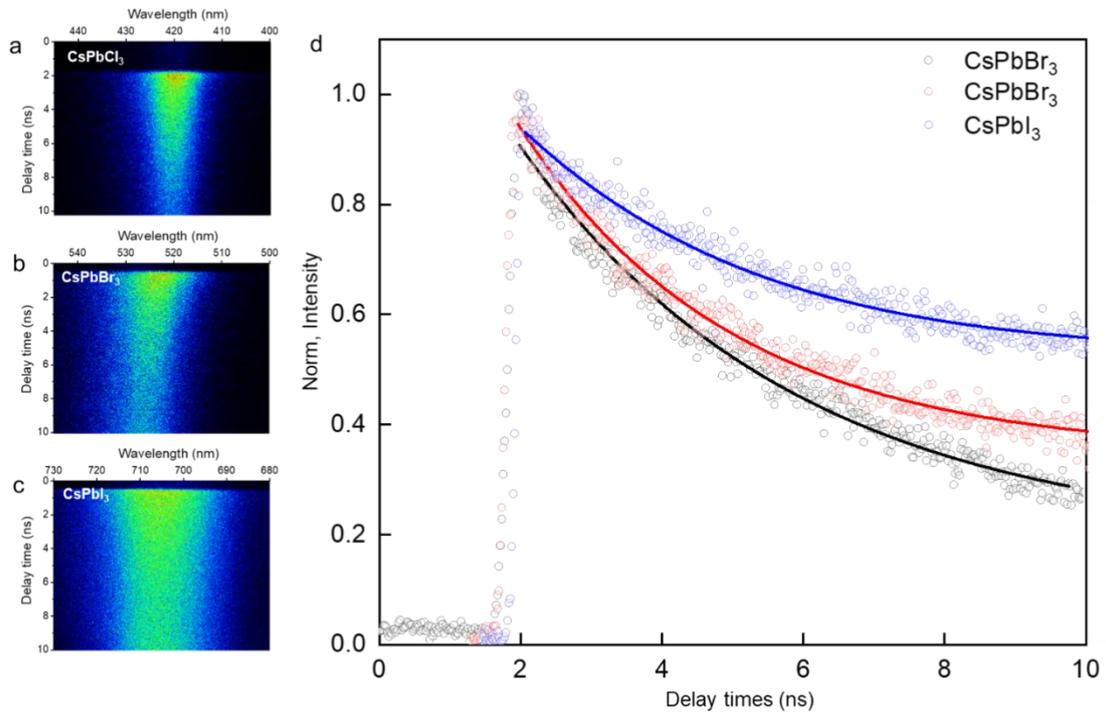

**Fig. S5 | Transient dynamic characterization of the representative CsPbX$_3$ MPs. a-c,** The steak camera images of CsPbCl$_3$, CsPbBr$_3$ and CsPbI$_3$ MPs. **d,** PL decay curves of CsPbCl$_3$, CsPbBr$_3$ and CsPbI$_3$ MPs.

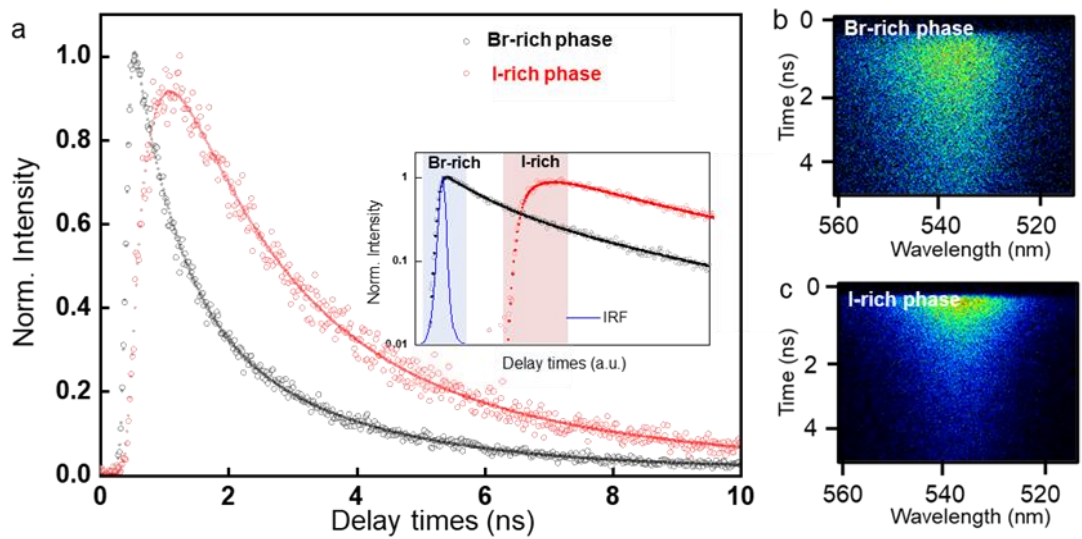

**Fig. S6 | Transient dynamic characterization of the representative Er-CsPb(Br$_x$I$_{1-x}$)$_3$ MP. a,** PL decay curves of Br-rich phase and I-rich phase perovskite in Er-CsPb(Br$_x$I$_{1-x}$)$_3$. The steak camera images of **b,** Br-rich phase perovskites and **c,** I-rich phase perovskites.

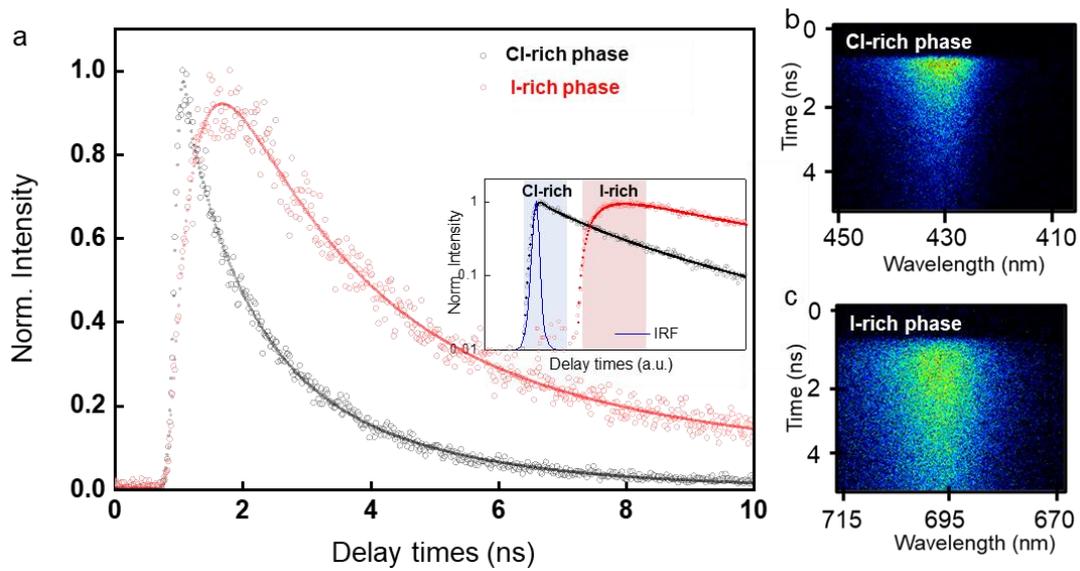

**Fig. S7 | Transient dynamic characterization of the representative Er-CsPb(Cl$_x$I$_{1-x}$)$_3$ MP. a,** PL decay curves of Cl-rich phase and I-rich phase perovskite in Er-CsPb(Cl$_x$I$_{1-x}$)$_3$. The steak camera images of **b,** Cl-rich phase perovskites and **c,** I-rich phase perovskites.

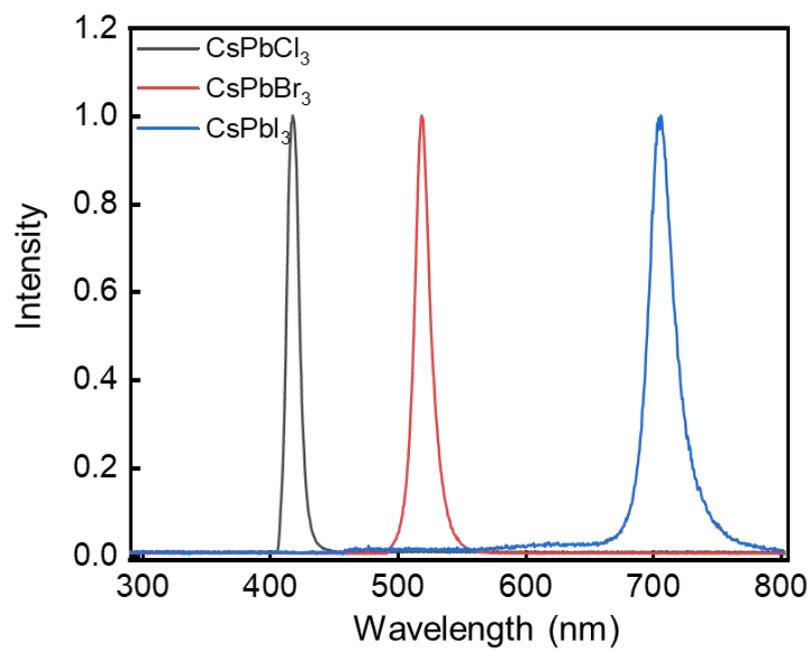

**Fig. S8 | Steady PL spectra of CsPbCl$_3$, CsPbBr$_3$ and CsPbI$_3$ MPs, respectively.**

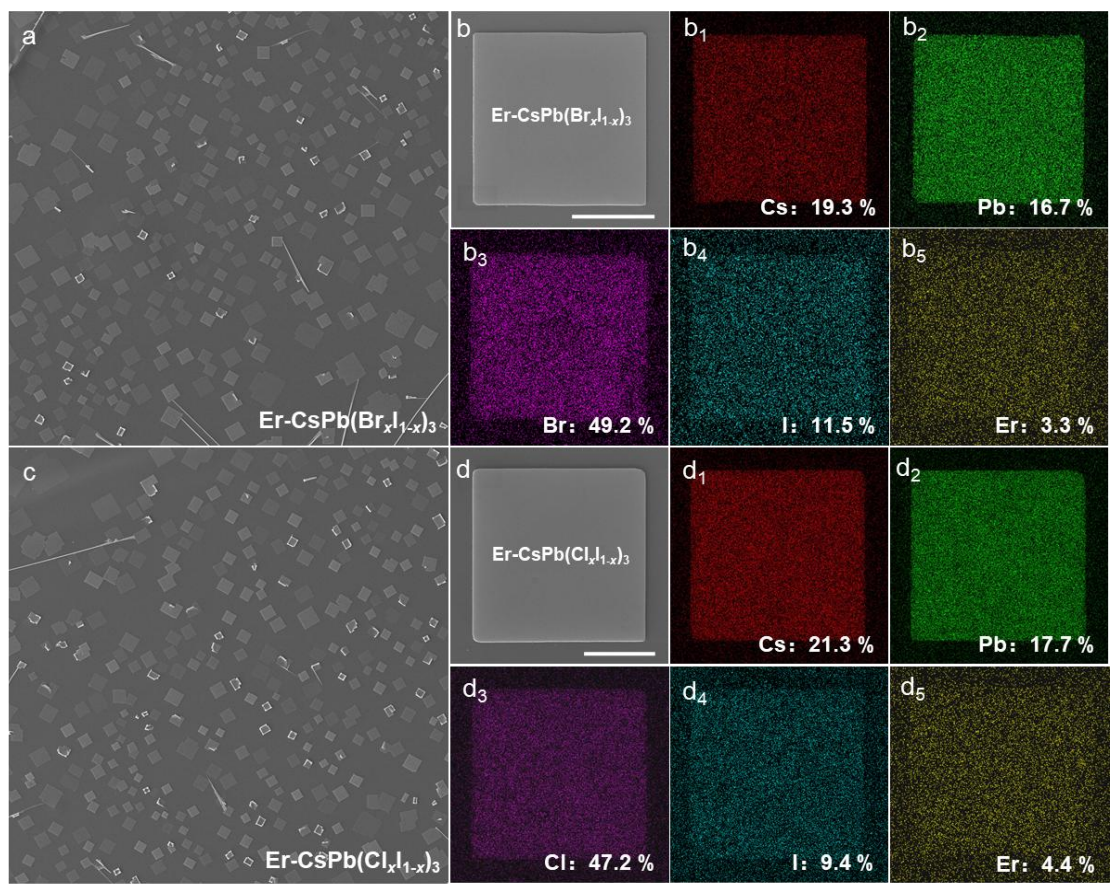

**Fig. S9 | Morphology of Er-CsPb(Br$_x$I$_{1-x}$)$_3$ and Er-CsPb(Cl$_x$I$_{1-x}$)$_3$ MPs. a,** Large-area low-magnification SEM image of typical Er-CsPb(Br$_x$I$_{1-x}$)$_3$ MPs. **b,** High-magnification SEM image of a selected Er-CsPb(Br$_x$I$_{1-x}$)$_3$ MP. Scale bar: 2 μm. **b$_1$-b$_5$,** Element mapping of the typical Er-CsPb(Br$_x$I$_{1-x}$)$_3$ MP. **c** Large-area low-magnification SEM image of typical Er-CsPb(Cl$_x$I$_{1-x}$)$_3$ MPs. **d,** High-magnification SEM image of a selected Er-CsPb(Cl$_x$I$_{1-x}$)$_3$ MP. Scale bar: 2 μm. **d$_1$-d$_5$,** Element mapping of the typical Er-CsPb(Cl$_x$I$_{1-x}$)$_3$ MP.

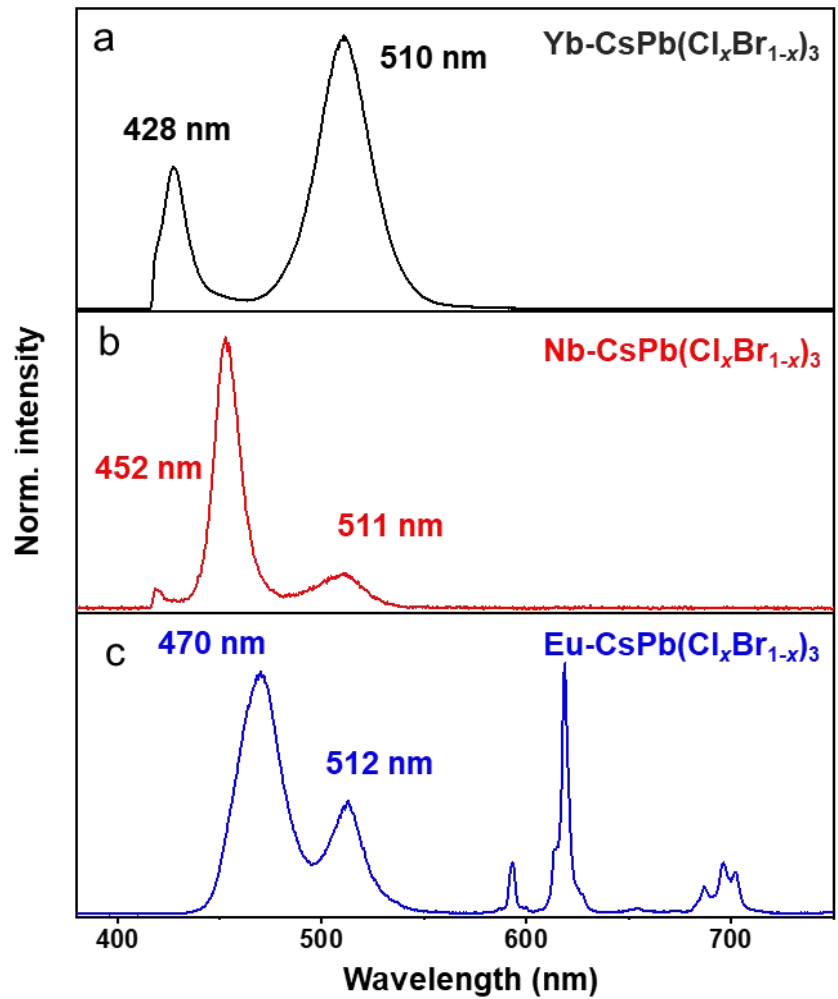

**Fig. S10 | Universality of the effect of lanthanide ion doping on spectra. a,** The PL spectrum of Yb-CsPb(Cl$_x$Br$_{1-x}$)$_3$. **b,** The PL spectrum of Nb-CsPb(Cl$_x$Br$_{1-x}$)$_3$. **c,** The PL spectrum of Eu-CsPb(Cl$_x$Br$_{1-x}$)$_3$.

**Methods for Fitting the Dynamic Curves**

The dynamic curves are fitted by a multi-exponential function with deconvolution of the IRF as follows:

$$N(t) \propto I(t) = \sum A_i \exp\left(-\frac{t}{\tau_i}\right) \qquad (1)$$

Where $N(t)$ is the amount of carrier at time $t$, and is proportional to the emission intensity $I(t)$. $A_i$ ($i = 1,2,3$) represents the weight percentage of the carrier contribution from each component. $\tau_i$ ($i = 1,2,3$) is the time constant of each component. Table S2 provides the fitting results for three pure-phase MPs. All of them exhibit two time constants of $\tau_1$ and $\tau_2$, which are attributed to the carrier recombination through surface state and intrinsic recombination. All of them exhibit two time constants, $\tau_1$ and $\tau_2$, which are attributed to carrier recombination through surface states and intrinsic recombination.

| Section | τ1(ns) | A1(%) | τ2(ns) | A2(%) | τ2(ns) | A2(%) |
|---|---|---|---|---|---|---|
| Cl-rich phase | 0.37 | 45.6 | 0.96 | 38.3 | 5.71 | 16.1 |
| Br-rich phase | 0.53 | -271.8 | 0.91 | 122.7 | 3.48 | 249.1 |

Table S1. The dynamic curve fitting data of Cl-rich phase and Br-rich phase perovskite of Er-CsPb(Cl$_x$Br$_{1-x}$)$_3$.

| Section | τ1(ns) | A1(%) | τ2(ns) | A2(%) |
|---------|--------|-------|--------|-------|
| $CsPbCl_3$ | 0.38 | 26 | 6.83 | 74 |
| $CsPbBr_3$ | 2.7 | 61 | 71.80 | 39 |
| $CsPbI_3$ | 2.48 | 39.75 | 0.6.40 | 60.25 |

**Table S2**. Dynamic Curve Fitting Data for $CsPbCl_3$, $CsPbBr_3$ and $CsPbI_3$.

| Section | τ1(ns) | A1(%) | τ2(ns) | A2(%) | τ2(ns) | A2(%) |
|---|---|---|---|---|---|---|
| Br-rich phase | 0.66 | 58.6 | 2.4 | 39.9 | 33.7 | 1.5 |
| I-rich phase | 0.45 | -149.6 | 1 | 152.2 | 3.8 | 97.4 |

**Table S3**. Dynamic Curve Fitting Data for Br-rich and I-rich Phases of Er-CsPb(Br$_x$I$_{1-x}$)$_3$.

| Section | τ1(ns) | A1(%) | τ2(ns) | A2(%) | τ2(ns) | A2(%) |
|---|---|---|---|---|---|---|
| Cl-rich phase | 0.02 | 76.4 | 0.9 | 13.3 | 2.56 | 10.3 |
| I-rich phase | 0.4 | -481 | 1.82 | 423 | 8 | 158 |

**Table S4**. Dynamic Curve Fitting Data for Cl-rich and I-rich Phases of Er-CsPb(Cl$_x$I$_{1-x}$)$_3$.

|          | Uniform distribution | Clustered distribution | Clustered minus uniform |
|----------|---------------------|------------------------|-------------------------|
| undoped  | -209.12             | -209.02                | +0.01                   |
| Er doped | -209.57             | -210.54                | -0.979                  |

**Table S5**. Formation energies of $CsPb(Cl_{0.83}Br_{0.17})_3$ with and without Er doping under different halide distributions. The energies are calculated for the uniform and clustered distributions of halides, both before and after $Er^{3+}$ doping.

|  | Intra-plane migration (eV) | Inter-plane migration (eV) |
| --- | --- | --- |
| CsPb(Cl$_{0.83}$Br$_{0.17}$)$_3$ | 0.56 | 1.00 |
| Er-CsPbCl$_3$ | 1.58 | 1.83 |

**Table S6**. Intra- and inter-octahedral migration barrier calculation of Cl ions in CsPb(Cl$_{0.83}$Br$_{0.17}$)$_3$ and Er-CsPbCl$_3$.

**A method for calculating the composition of mixed perovskites based on PL spectra**

The band gap of $CsPb(X_xY_{1-x})_3$ (X and Y represent halide ions including Cl, Br, and I) alloy perovskites can be determined by the following formula:

$$E_{g(x)} = xE_{g1} + (1-x)E_{g2} - mx(1-x) \qquad (1)$$

$E_{g1}$ and $E_{g2}$ are the bandgaps of the corresponding one-component halide perovskite, here corresponding to $CsPbCl_3$ 2.9eV, $CsPbBr_3$ 2.33 eV and $CsPbI_3$ 1.7 eV, respectively. m is the band bending parameter, here we value 0.48.